\documentclass{aa}
\usepackage{graphicx,natbib}
\bibpunct{(}{)}{;}{a}{}{,}
\usepackage{txfonts}

\begin{document}

\title{Planet gaps in the dust layer of 3D protoplanetary disks: }
\subtitle{I. Hydrodynamical simulations of T Tauri disks}
\titlerunning{Planet gaps in the dust layer of 3D protoplanetary disks. I}
\author{L. Fouchet
       \inst{1}
       \and
         J.-F. Gonzalez \inst{2}
       \and
       S. T. Maddison \inst{3}}
\offprints{L. Fouchet}
\institute{Physikalisches Institut, Universit\"at Bern, CH-3012 Bern, Switzerland\\
        \email{laure.fouchet@space.unibe.ch}
        \and
        Universit{\'e} de Lyon, Lyon, F-69003, France; Universit\'e Lyon 1, Villeurbanne, F-69622, France; CNRS, UMR 5574, Centre de Recherche Astrophysique de Lyon; \'Ecole Normale Sup\'erieure de Lyon, 46 all\'ee d'Italie, F-69364 Lyon cedex 07, France\\
        \email{Jean-Francois.Gonzalez@ens-lyon.fr}
	\and
        Centre for Astrophysics and Supercomputing, Swinburne University, PO Box 218, Hawthorn, VIC 3122, Australia \\
        \email{smaddison@swin.edu.au}
}

\date{Received 30 November 2009 / Accepted 7 May 2010}

\abstract
{While sub-micron- and micron-sized dust grains are generally well mixed with the gas  phase in protoplanetary disks, larger grains will be partially decoupled and as a consequence have a different distribution from that of the gas.  This has ramifications for predictions of the observability of protoplanetary disks, for which gas-only studies will provide an inaccurate picture. Specifically, criteria for gap opening in the presence of a planet have generally been studied for the gas phase, whereas the situation can be quite different in the dust layer once grains reach mm sizes, which is what will be observed by ALMA.}
{We aim to investigate the formation and structure of a planetary gap in the dust layer of a protoplanetary disk with an embedded planet.}
{We perform 3D, gas$+$dust SPH simulations of a protoplanetary disk with a planet on a fixed circular orbit at 40 AU to study the evolution of both the gas and dust distributions and densities in the disk. We run a series of simulations in which the planet mass and the dust grain size varies.}
{We show that the gap in the dust layer is more striking than in the gas phase and that it is deeper and wider for more massive planets as well as for larger grains. For a massive enough planet, we note that cm-sized grains remain inside the gap in corotation and that their population in the outer disk shows an asymmetric structure, a signature of disk-planet interactions even for a circular planetary orbit, which should be observable with ALMA.}
{}

\maketitle

\section{Introduction}
\label{sec:introduction}

While we understand the general scenario of planet formation via accretion \citep{Mizuno1980, Wetherill1980, Lissauer1987, Wetherill1990, Pollack_etal1996}, the devil is in the detail and the processes by which tiny sub-micron grains grow into planetesimals, the building blocks of planets, is not well understood. With over 400 extrasolar planets detected to date as listed in the extrasolar planet encyclopedia\footnote{http://exoplanet.eu/}, we can start to do some meaningful statistics on the types of planets that exist in our galaxy \citep{AMBW05,M07} to help constrain theories of planet formation.

A wealth of analytical as well as numerical studies of the formation of gaps by a planet in a gas disk have shown how the shape of the gap depends on properties of the planet (mass) as well as the gaseous disk (pressure scale height, viscosity). See \cite{CMM06} for the case of gap shapes and \cite{PNKMA07} for a more general review on planet migration and gap formation.

The dust phase, however, has been shown to behave differently to the gas. Dust experiences a headwind from the pressure-supported sub-Keplerian gas and the induced drag force slows the dust and makes it settle to the midplane and migrate inwards. The magnitude of these effects depends strongly on the grain size and the disk density \citep{W77,SV96,SV97,GBFL04,GL04,BF05}.

In recent years, gap formation by a planet embedded in disks of gas and dust has been studied by several authors, for different planet masses and different sizes of the solid particles. \citet{PM04,PM06} determined the spatial distribution of 1~mm grains with 2D simulations in order to derive the smallest planet mass that would result in an observable gap with ALMA (Atacama Large Millimeter Array). In their work the dust was strongly coupled to the gas and responded indirectly to the planet gravity through the radial pressure gradients caused by a Neptune-mass planet in the gas disk that led to the formation of a gap. \citet{Muto09} investigated the effect of a low-mass planet on the dust distribution by injecting one dust grain at a time and derived criteria for gap opening. \citet{Ciecielag07} as well as \citet{Marzari00} focused on already formed planetesimals while the gas phase is still present. \citet{Ciecielag07} considered planetesimals down to 1~m in size. They studied the effect of spiral structures in the circumprimary gas disk triggered by a secondary companion in a tight binary system in order to derive relative velocities between planetesimals and determine whether planet formation is possible in such systems.

The presence of planets, and the gap they create when they are massive enough, can also help constrain the global properties of the gas (temperature, density, viscosity) as well as those of the dust (grain size distribution, degree of settling) in the disks. This can be achieved by measuring the width and brightness of the gap, ideally for each phase \citep{W05,CMM06,Fouchet07}.

It has been shown that ALMA will be able to observe a planetary gap opened by a 1~$M_J$ planet at 5.2~AU from a 1~$M_\odot$ star at a distance of 140~pc under the naive assumption that gas and dust are well mixed \citep{W02}. It will certainly be possible to observe other features related to the presence of the planet, such as warm dust in its vicinity or even spiral waves for distances not exceeding 100~pc \citep{W05}. Here we focus on the formation and features of the gap itself in the case of a massive protoplanet.

Our previous simulations of dust evolution in a typical Classical T-Tauri Star (CTTS) disk \citep[hereafter BF05]{BF05}\defcitealias{BF05}{BF05} showed that the thickness of the dust layer depends on grain size because different sized grains fall to the midplane at different rates. We distinguished three dynamical regimes for the dust: (1) almost uncoupled for large grains where the dust component follows slightly perturbed Keplerian orbits and keeps its 3D distribution (if initially 3D); (2) weakly coupled for intermediate-sized grains for which settling is very efficient; (3) strongly coupled for small grains where grains are forced to follow the gas motion.

We have also investigated the formation of a gap by a planet immersed in a Minimum Mass Solar Nebula (MMSN) and showed that dust settling makes the gap much more striking in the dust layer than in the gas phase because of its reduced vertical extension for weakly coupled grains \citep[hereafter F07]{Maddison07,Fouchet07}\defcitealias{Fouchet07}{F07}. Indeed, the criterion for gap formation depends on the disk scale height \citep{CMM06}. Because of the size of particles in the weakly coupled regime for the particular case of the MMSN (1~m, see Sect.~\ref{sec:theory}), that study had no direct application to observations. Images of dusty disks at infrared wavelengths do not probe such large particles, but instead trace the smaller, strongly coupled dust grains whose distribution is similar to that of the gas.

In this paper, we examine the gap formation in the dust layer of CTTS disks, which are spatially more extended and less dense than the theoretical MMSN case. Our ultimate goal is to use the results of our hydrodynamical simulations to produce synthetic images for ALMA. We consider in particular the effects of a massive planet in the outer cooler regions of the disk. Indeed, several planets at large distances from their star have been detected, such as those recently announced orbiting Fomalhaut \citep{Kalas08}, or HR~8799 \citep{Marois08}. The extrasolar planet encyclopedia lists 10 planets with a semi-major axis larger than 20~AU, and 7 of them have minimum masses larger than 5~$M_\mathrm{J}$. We study the dust distribution in the weakly coupled regime, which corresponds for these CTTS disks to a size range (100 $\mu$m to 1 cm) that can directly be probed by current and future (sub)millimetre instruments.

The present paper is the first part of this work, presenting the hydrodynamical simulations. In Sect.~\ref{sec:theory}, we discuss the gas-dust interaction in the presence of a planet. In Sect.~\ref{sec:simulations}, we describe the numerical method and simulation suite. Results are described in Sect.~\ref{sec:results}, while the analysis and explanations are presented in Sect.~\ref{sec:discussion}. We conclude in Sect.~\ref{sec:conclusion}. In a forthcoming companion paper, we will use the simulations presented here to produce synthetic images for ALMA and present the most favorable observing configurations to detect the gap and associated structures.

\section{Dust behaviour under the effect of gas drag}
\label{sec:theory}

Dust grains immersed in a gas disk experience a drag force which, in the Epstein regime \citepalias[valid for the nebula parameters and grain sizes we consider, see ][]{BF05}, is given by
\begin{equation}
\vec{F}_\mathrm{D}=\displaystyle\frac{m_\mathrm{d}}{t_\mathrm{s}}\,\Delta\vec{v},
\label{EqDragForce}
\end{equation}
where
\begin{equation}
t_\mathrm{s}=\displaystyle\frac{\rho_\mathrm{d}\,s}{\rho_\mathrm{g}\,c_\mathrm{s}}
\label{EqStoppingTime}
\end{equation}
and $m_\mathrm{d}$ is the grain mass, $t_\mathrm{s}$ the stopping time, $\Delta\vec{v}=\vec{v}_\mathrm{d}-\vec{v}_\mathrm{g}$ the differential gas-dust velocity, $\rho_\mathrm{d}$ the dust intrinsic density, $s$ the grain size, $\rho_\mathrm{g}$ the gas density and $c_\mathrm{s}$ the gas sound speed.

Early studies by \citet{W77} showed that there exists a grain size for which radial migration is fastest, satisfying the condition $t_\mathrm{s}\Omega_\mathrm{K}=1$ where $\Omega_\mathrm{K}$ is the Keplerian pulsation. This optimal grain size is thus given by
\begin{equation}
s_\mathrm{opt}=\frac{\rho_\mathrm{g}\,c_\mathrm{s}}{\rho_\mathrm{d}\,\Omega_\mathrm{K}}.
\label{EqSopt}
\end{equation}
This depends on the nebula parameters, and in particular on the gas density profile. For a vertically isothermal disk in hydrostatic equilibrium, the gas density profile is $\rho_\mathrm{g}(r,z)=\rho_\mathrm{g}(r,0)\,e^{-z^2/2H(r)^2}$, where the disk scale height is $H(r)=c_\mathrm{s}(r)/\Omega_\mathrm{K}(r)$. Integrating vertically yields the surface density $\Sigma_\mathrm{g}(r)=\sqrt{2\pi}\,\rho_\mathrm{g}(r,0)\,H(r)$. Consequently, the optimal size in the disk midplane is
\begin{equation}
s_\mathrm{opt}(r,0)=\frac{\Sigma_\mathrm{g}(r)}{\sqrt{2\pi}\,\rho_\mathrm{d}}.
\label{EqSopt2}
\end{equation}
This is typically $\sim$1~mm--1~cm for CTTS disks \citepalias{BF05} and $\sim$1~m for MMSN disks (\citealt{W77}; \citetalias{Fouchet07}). The three regimes mentioned in Sect.~\ref{sec:introduction} correspond to grains much larger than $s_\mathrm{opt}$ (almost decoupled from the gas), of comparable size to $s_\mathrm{opt}$ (weakly coupled), and much smaller than $s_\mathrm{opt}$ (strongly coupled). The intermediate regime, for which both radial migration and vertical settling are efficient, is the most dynamically interesting.

\citet{HB03} have shown that solid particles drift towards pressure maxima. Indeed, at the inner edge of a pressure maximum, pressure {increases with radius} and the pressure gradient is positive, acting in the same direction as the stellar gravity. Gas has to flow super-Keplerian in order not to fall on the star. It speeds up the dust that, with no internal pressure, would flow at Keplerian velocities. In order to conserve angular momentum, which increases with radius in a Keplerian disk, dust that is sped up and therefore gains angular momentum needs to migrate outwards. At the outer edge, this is all reversed and dust, slowed down by a sub-Keplerian gas, has to drift inwards. As a result, dust is concentrated in the pressure maximum.

From those considerations, in a disk with an embedded planet we therefore expect dust grains to migrate towards the pressure maxima on either side of the gap and to accumulate there. This behaviour has been observed by, e.g., \citet{PM04,PM06}, \citet{Maddison07}, and \citetalias{Fouchet07}.

\section{Simulations}
\label{sec:simulations}

\subsection{Code description}
\label{sec:code}

We have developed a 3D, two-phase (gas$+$dust) Smoothed Particles Hydrodynamic (SPH) code \citepalias{BF05}. We use it to model a protoplanetary disk of mass $M_\mathrm{disk}$, orbiting a star of mass $M_\star$, and containing a planet of mass $M_\mathrm{p}$ on a fixed circular orbit of radius $r_\mathrm{p}$. The code units are chosen such that $G=M_\star=r_\mathrm{p}=1$. The disk is treated as vertically isothermal, implying that the cooling is very efficient. Any heat produced by viscous dissipation or by stellar irradiation is radiated away much faster than the gas dynamical timescale. Disk self-gravity is not implemented: it is negligible for the low-mass disks we study. Gas and dust are treated as two inter-penetrating fluids coupled by gas drag in the Epstein regime. In each simulation we consider a population of uniform-sized grains which do not shatter or grow. For full details of the code, see \citetalias{BF05}.

We use the  standard SPH viscosity \citep{JJM89} with $\alpha_\mathrm{SPH}=0.1$ and $\beta_\mathrm{SPH}=0.5$. The $\alpha_\mathrm{SPH}$ term was introduced to remove subsonic velocity oscillations that follow shocks \citep{Monaghan83} and the $\beta_\mathrm{SPH}$ term damps high Mach number shocks and prevents particle interpenetration \citep{JJM89}. The $\alpha_\mathrm{SPH}$ was shown by \citet{Monaghan85} to produce shear and bulk viscosity. Our choice of artificial SPH viscosity terms leads to a Shakura-Sunayaev $\alpha_\mathrm{SS}$ parameter for the viscosity of order $10^{-2}$. This is within the range of values indicated by observations of protoplanetary disks \citep{Hartmann1998,King2007}. For a discussion on the validity of these viscosity parameters in protoplanetary disks, see \citetalias{Fouchet07}. In that previous work, we used $\alpha_\mathrm{SPH}=0.1$ and $\beta_\mathrm{SPH}=0$ and showed that the resulting $\alpha_\mathrm{SS}$ was only slightly smaller than with $\beta_\mathrm{SPH}=0.5$. In this work, we consider a more massive planet, which will generate a stronger spiral density wave. As a result, we need to increase $\beta_\mathrm{SPH}$.

SPH has clear advantages over grid-based codes, in particular when investigating boundary-free problems such as 3D flared protoplanetary disks. However, within the simulation community, SPH is often considered to be a poor choice when modelling planet-disk interactions since the resolution decreases in the gap because of the reduced number of particles there. \citet{deValBorro06} compared the results from grid-based and SPH codes when modelling the planet-disk interactions and found that SPH codes predict the correct shape of the gap, albeit with less resolution in the low density regions and with weaker planetary wakes.  It should be noted, however, that the disk model of the grid-based codes of \citet{deValBorro06} used a rather low viscosity (which SPH cannot achieve). If one were instead to consider more turbulent disks, as expected in nature, that gap would be shallower and edges not as sharp in both SPH and grid-based codes. Authors modeling low-viscosity regions (e.g. dead zones) expect a gap with razor-sharp edges, while it is certainly not the case for the more viscous disks we consider. As mentioned above, we use $\alpha_\mathrm{SS}\simeq10^{-2}$ as suggested by observations of disks.

While the SPH community is substantially smaller than the grid-based community, people have been trying to improve the SPH technique by including high order algorithms in Lagrangian methods. \citet{Inutsuka02} proposes a new formalism called GSPH, for Godunov Smooth Particle Hydrodynamics, which uses a Riemann solver to improve the treatment of shocks. One issue with this approach is that complicated equations of state are more difficult to implement than when one relies on artificial viscosity \citep[see][]{M97}. \citet{Maron03} propose Gradient Particle Magnetohydrodynamics, but, as discussed by \citet{Price04}, it is not clear whether the increased complexity and computing cost is compensated by the gain in accuracy. \citet{Borve09} propose Regularized Smooth Particle Hydrodynamics where the solution is mapped onto a regular grid. This approach reduces the numerical noise but increases the diffusivity. A novel technique has recently been presented by \citet{Springel10}, where a moving grid is set up by means of tessellation. This technique combines the advantages of both Eulerian and Lagrangian techniques and seems very promising. These developments are mostly proofs of concepts and need to be extensively tested. For this work we continue with the standard SPH whose achievements and limitations have been clearly addressed in the literature.

\subsection{Simulation parameters}
\label{sec:sim-param}

We model a typical CTTS disk with $M_\star=1~M_{\sun}$, $M_\mathrm{disk}=0.02~M_{\sun}$ and comprising 1\% dust by mass. The disk initially extends from 0.1 to 3 code units, which corresponds to 4--120~AU, with a planet at orbital radius $r_\mathrm{p}=40$~AU =~1~code unit. The initial surface density profile, $\Sigma(r)=\Sigma_0 r^{-p}$, is taken to be flat ($p=0$) to follow our previous work \citepalias{Fouchet07}. Initially, $s_\mathrm{opt}$ is therefore also uniform in radius, with a value of $\sim$1.5~cm in the midplane. The initial temperature profile is $T(r)=T_0 r^{-q}$, with $q=1$. The disk aspect ratio is initially chosen to be $H/r=0.05$ at $r_\mathrm{p}$. The intrinsic grain density, $\rho_\mathrm{d}$, is 1,000~kg\,m$^{-3}$.

As in \citetalias{Fouchet07}, we embed the planet in a gas disk and evolve the system for 8 planetary orbits and then add the dust phase. The star is kept fixed at the origin. The evolution is then followed for a total of 104 planetary orbits. The disk is allowed to spread outwards and particles are only removed from the simulation if they go beyond 4~code units (160~AU). The inner radius is fixed at 0.1~code units (4~AU) and particles crossing that limit are assumed to be accreted by the star. \citet{Crida07} have shown that the disk interior to the gap is dramatically depleted if the inner boundary condition is too close to the planet. Our test simulations show that an open inner boundary at 0.1~$r_\mathrm{p}$ avoids this depletion of the inner gas disk, as extensively discussed in \citetalias{Fouchet07} (see their Fig.~11). For full details of the initial setup, see \citetalias{Fouchet07}. All simulations contain 200,000 particles per phase.

\begin{table}
\begin{center}
\caption{Simulation suite.}
\label{tab-simparams}
\begin{tabular}{r@{~}lccc}
\hline
                &                    & $s=100~\mu$m & $s=1$~mm & $s=1$~cm \\
\hline
$M_\mathrm{p}=$ & $0.1~M_\mathrm{J}$ &              & $\bullet$ & $\bullet$ \\
$M_\mathrm{p}=$ & $0.5~M_\mathrm{J}$ &              & $\bullet$ & $\bullet$ \\
$M_\mathrm{p}=$ & $1~M_\mathrm{J}$   &              & $\bullet$ & $\bullet$ \\
$M_\mathrm{p}=$ & $5~M_\mathrm{J}$   & $\bullet$    & $\bullet$ & $\bullet$  \\
\hline
\end{tabular}
\end{center}
\end{table}

We run a series of simulations in which we vary the grain size $s$ from 100~$\mu$m to 1~cm, so that dust is in the weakly coupled regime, and the planet mass, $M_\mathrm{p}$, from 0.1 to 5~$M_\mathrm{J}$, in order to study how these parameters affect the resulting dust distribution around the planetary gap. Here we investigate larger planet-to-star mass ratios than in \citetalias{Fouchet07}, which produce a stronger spiral density wave. Table~\ref{tab-simparams} presents the simulation suite.

\section{Results}
\label{sec:results}

In this section, we describe the results of the simulations. Analysis and discussion of the results will be presented in the next section.

\begin{figure*}[t]
\sidecaption
\includegraphics[width=12cm]{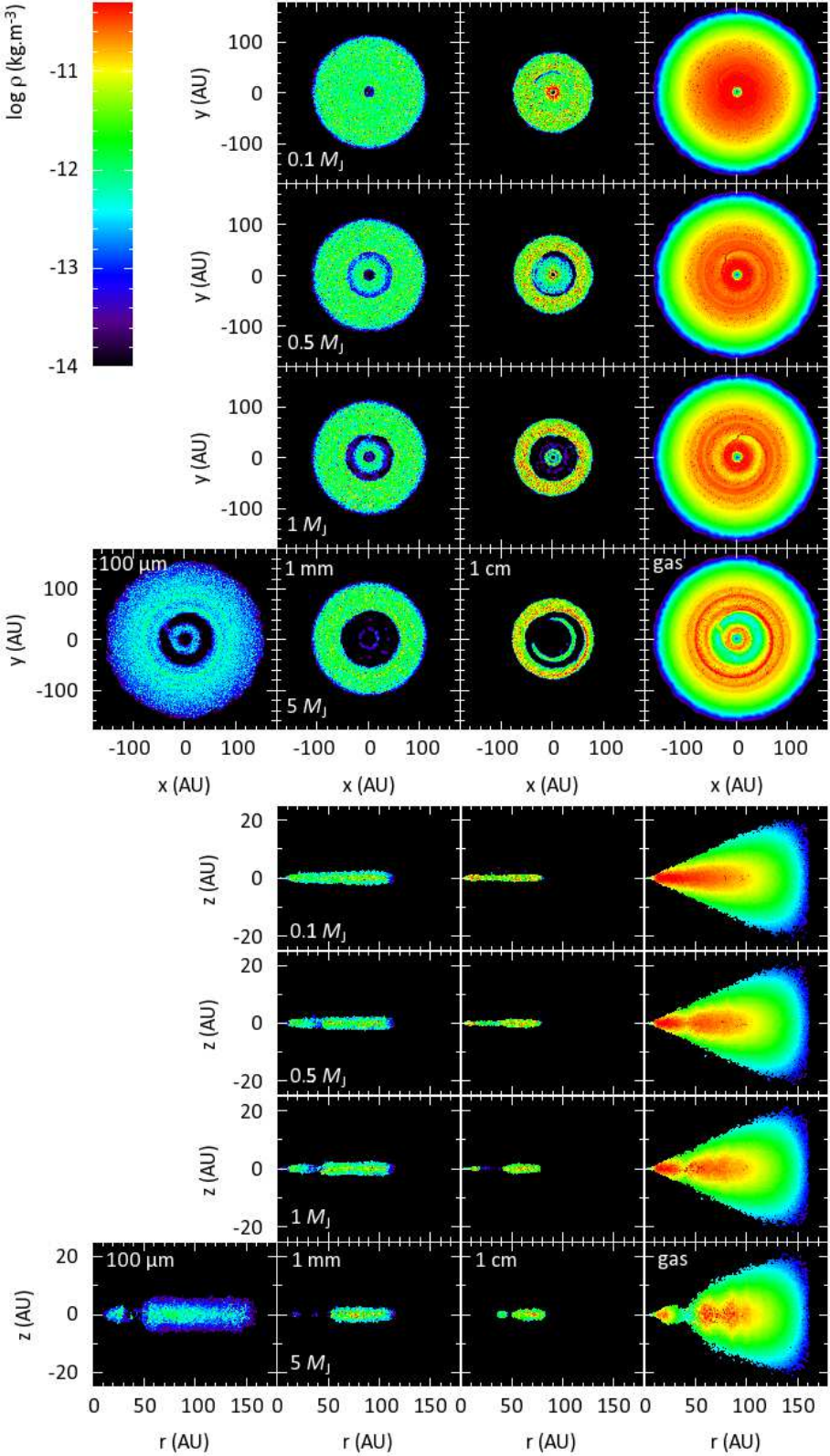}
\caption{Density maps in midplane (top) and meridian plane (bottom) cuts of the disks. The three leftmost columns show the dust density for $s=100~\mu$m, 1~mm and 1~cm, from left to right, and the right hand column shows the gas density. The rows show simulations with $M_\mathrm{p}=0.1$, 0.5, 1 and $5~M_\mathrm{J}$, from top to bottom.}
\label{density_maps}
\end{figure*}

\begin{figure*}[t]
\sidecaption
\includegraphics[width=12cm]{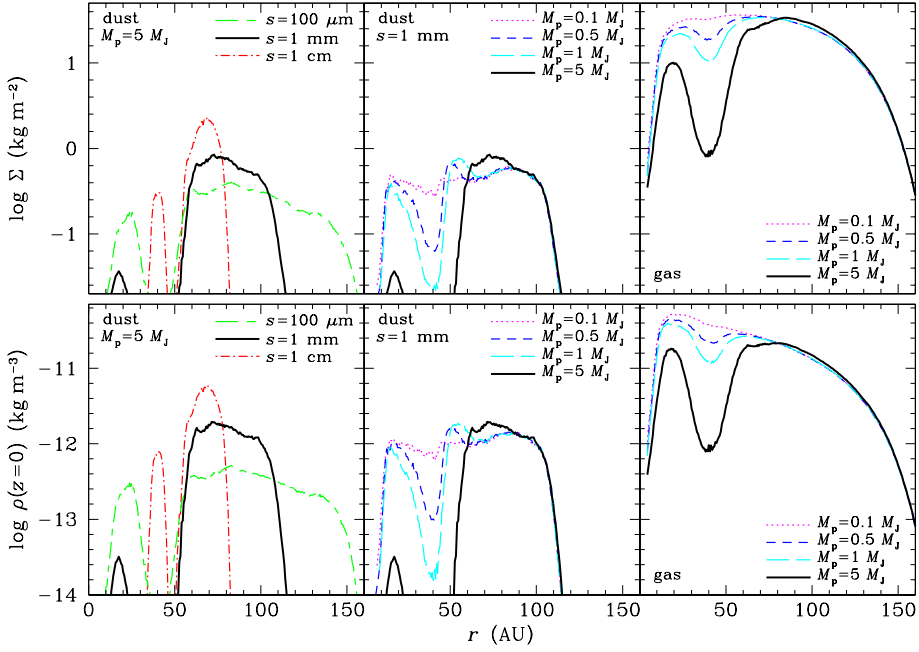}
\caption{Azimuthally averaged surface density (top) and midplane volume density (bottom) profiles after 104 planetary orbits. Left: dust densities for $M_\mathrm{p}=5~M_\mathrm{J}$ and different grain sizes. Center: dust densities for $s=1$~mm and different planet masses. Right: gas densities for $s=1$~mm and different planet masses.}
\label{density_profiles}
\end{figure*}

In Fig.~\ref{density_maps} we plot the dust and gas distributions in the $xy$ and $rz$ planes (where the center of frame is fixed on the star) after 104 planet orbits, coloured by volume density for all simulations. Fig.~\ref{density_profiles} shows the azimuthally averaged radial surface density profile and the midplane volume density of the dust for disks with a 5~$M_\mathrm{J}$ planet and grains sizes from 100~$\mu$m to 1~cm, and of the gas and dust for disks with $s=1$~mm and planet masses from 0.1 to 5~$M_\mathrm{J}$.

The gas is little affected by the dust phase (see Sect.~\ref{sec:BR}) and its distribution is similar for all grain sizes. Its density is shown in the rightmost column of Fig.~\ref{density_maps} and the right panels of Fig.~\ref{density_profiles} for $s=1$~mm and all planet masses. As  expected (see Sect.~\ref{sec:introduction}), the more massive the planet, the deeper the gap. We find that a 5~$M_\mathrm{J}$ planet carves a well-defined gap in the gas component of our CTTS disk, that planets with masses of 0.5 or 1~$M_\mathrm{J}$ create shallow gaps, and that a 0.1~$M_\mathrm{J}$ planet only slightly perturbs the gas density profile.

The dust phase, however, has a very different distribution depending on the grain size, as can be seen by comparing all simulations for $M_\mathrm{p}=5~M_\mathrm{J}$ on the bottom rows of Fig.~\ref{density_maps} and the left panels of Fig.~\ref{density_profiles}. The disk extension is dramatically reduced as $s$ increases from 100~$\mu$m to 1~cm: both its outer radius and vertical thickness decrease. A spiral density wave is visible on the face-on views for 100~$\mu$m and 1~mm grains (Fig.~\ref{density_maps}), similar to that seen in the gas disk, while only a small section of it is visible in the much narrower disk of 1~cm grains. The gap is wider and deeper, with higher densities at its outer edge, for larger grain sizes, and in all cases more pronounced than in the gas phase. For 1~cm grains, because they are the most efficiently settled, the midplane volume density better shows their concentration at the gap outer edge, with a higher dust-to-gas ratio ($\sim0.3$) than the surface density does ($\sim0.09$, compared to the initial uniform value of 0.01). The disk interior to the gap is virtually unaffected by the presence of the planet for 100~$\mu$m grains, but has almost disappeared for the 1~mm grains and is no longer present for 1~cm grains, for which a population of grains in corotation with the planet can be seen instead. In that latter case, the gap is slightly asymmetric and the outer disk appears eccentric (Fig.~\ref{density_maps}).

The same effect of grain size is seen for other planet masses in the three upper rows of Fig.~\ref{density_maps}: the dust disk's outer radius and vertical extension are smaller, whereas the planet's effect is larger, for 1~cm grains than for 1~mm grains.

The effect of the planet mass on the dust phase for those two sizes can be observed in the two center columns of Fig.~\ref{density_maps} and for 1~mm grains in the center panels of Fig.~\ref{density_profiles}. The disk has the same radial extent but is slightly thinner for less massive planets. As is seen for the gas, the gap width and depth increase as $M_\mathrm{p}$ increases, and the gap is always more pronounced in the dust phase than in the gas. A very shallow gap is visible in the dust for a 0.1~$M_\mathrm{J}$ planet, where only a slight perturbation is seen in the gas, and planets of 0.5 or 1~$M_\mathrm{J}$ already carve well-defined gaps in the dust disk. Contrary to the 5~$M_\mathrm{J}$ case, the disk interior to the gap is still present and particles in corotation are not observed.

\begin{figure*}[t]
\sidecaption
\includegraphics[width=12cm]{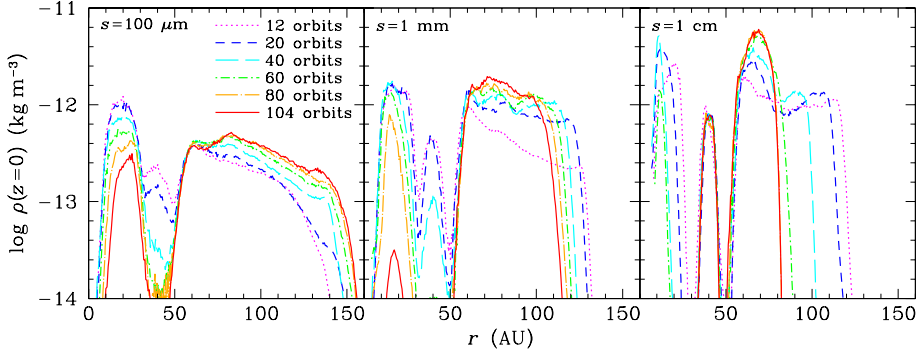}
\caption{Time evolution of the azimuthally averaged midplane radial volume density profiles for $M_\mathrm{p}=5~M_\mathrm{J}$ and different grain sizes.}
\label{density_profiles_time}
\end{figure*}

Figure~\ref{density_profiles_time} plots the time evolution of the azimuthally averaged surface density radial profiles for the three grain sizes for a 5~$M_\mathrm{J}$ planet. It shows that the disk's outer radius changes very little for 100~$\mu$m grains, decreases slowly for 1~mm grains, and very rapidly for 1~cm grains. The same behaviour can be seen for the density in the disk interior to the gap. For 1~cm grains, the displacement of the density peak towards the center indicates that the whole inner disk is accreted by the star. The gap formation shows the opposite trend as it is depleted rather rapidly for 100~$\mu$m grains, more slowly for 1~mm grains, and the density peak at the planet's orbital radius indicating the 1~cm corotating grains does not evolve.

Recently it has been shown that material can become trapped in the Lagrange points in disks with planets (e.g. in the dust-only+planet disks of \citet{Wolf2007} and in the gas+planet disks of \citet{deValBorro06}). Our simulations do not take the stellar wobble into account and thus cannot model Lagrangian points and produce any accumulation of matter there. Be that as it may, while gas is seen to be trapped at L4 and L5 in the inviscid simulations of de Val-Borro et al. (2006), viscosity eventually removes gas from these points and indeed, their viscous runs do not show such features. We therefore do not expect to see accumulation of either gas or dust at L4 and L5 in our viscous disks.

\section{Discussion}
\label{sec:discussion}

\subsection{Understanding the gap}
\label{sec:gap}

\begin{figure}[t]
\resizebox{\hsize}{!}{\includegraphics[angle=-90]{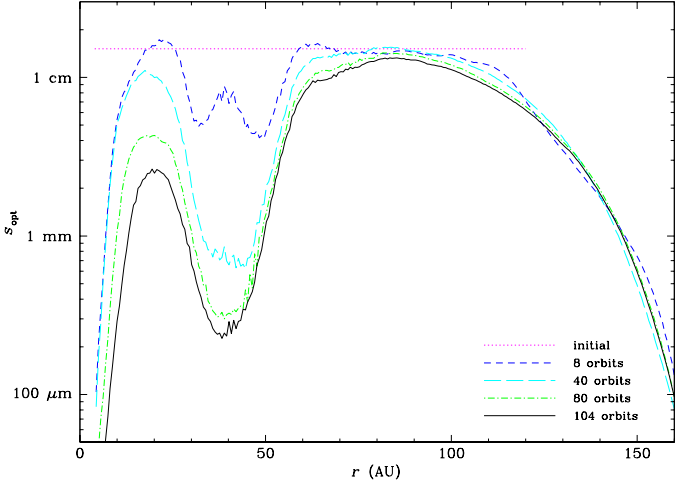}}
\caption{Evolution of the radial profile of the optimal grain size, $s_\mathrm{opt}$, in the midplane for $M_\mathrm{p}=5~M_\mathrm{J}$.}
\label{fig_sopt}
\end{figure}

Most of the features described in Sect.~\ref{sec:results} can be explained by comparing the grain size to the optimal size $s_\mathrm{opt}$, whose radial profile in the midplane is shown in Fig.~\ref{fig_sopt} at different times for $M_\mathrm{p}=5~M_\mathrm{J}$. The closer their size is to $s_\mathrm{opt}$, the more efficiently dust grains settle to the midplane and radially migrate.

Initially at a uniform value of 1.5~cm, $s_\mathrm{opt}$ decreases only slightly outside the gap and stays of the order of 1~cm out to $\sim$120~AU. Centimetre-sized grains therefore migrate and settle very rapidly, resulting in a compact and thin disk seen in Fig.~\ref{density_maps}. A similar behaviour is seen in the inner disk, which is very rapidly emptied as particles accrete onto the star, before $s_\mathrm{opt}$ decreases significantly below 1~cm there (Figs.~\ref{density_profiles_time} and \ref{fig_sopt}). In the gap, however, $s_\mathrm{opt}$ drops very rapidly from its value of $\sim1$~cm at dust injection (8~orbits) to below 1~mm in just 40~orbits, due to the rapid carving of the gap in the gas disk. Centimetre-sized grains are thus quickly decoupled from the gas and stay on their initial orbits in the gap, in corotation with the planet. Millimetre-sized grains settle and migrate less efficiently in the outer disk since their size is farther from $s_\mathrm{opt}$. In the inner disk, their distribution evolves slowly at first, when $s_\mathrm{opt}$ is still close to 1~cm, then their depletion accelerates as $s_\mathrm{opt}$ decreases and gets closer to their size, until only a tenuous annulus of dust is left at the end of the simulation. We anticipate it will disappear altogether at later times. In the gap, $s_\mathrm{opt}$ drops rapidly to 1~mm and below, resulting in the fast migration of 1~mm-sized grains out of the gap. Finally, 100~$\mu$m grains are more strongly coupled to the gas and evolve slowly on either side of the gap, since their size stays well below $s_\mathrm{opt}$ at all times. The increase of the midplane density in the outer disk reflects their moderate settling. In the gap, on the other hand, $s_\mathrm{opt}$ quickly approaches 100~$\mu$m, causing their efficient depletion.

We note that the thickness of the dust layer is not vanishingly small despite the lack of explicit turbulence. Indeed, the final shape of the dust layer in these simulations, in both radial extent and thickness, depends on the grain size. This is due to a complex interplay between both settling and radial migration times and grain size. The disk can be thicker in places due to radial migration pushing more material inwards but a change in $s_\mathrm{opt}$ (due to density) further inwards can cause the dust to pile up. This dictates the shape of the dust disk during the disk evolution. As it is a highly non-linear process, it is difficult to predict the final shape of the dust layer.

\begin{figure}[t]
\resizebox{\hsize}{!}{\includegraphics[angle=-90]{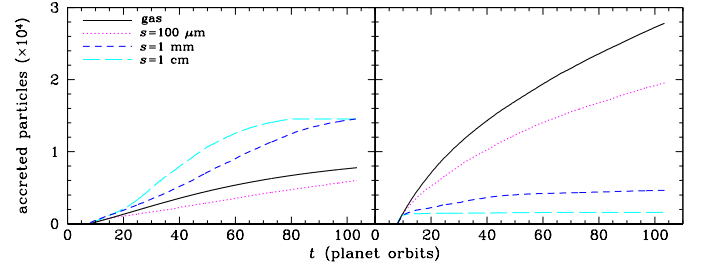}}
\caption{Number of SPH particles accreted by the star (left) and the planet (right) as a function of time, for $M_\mathrm{p}=5~M_\mathrm{J}$.}
\label{accretion}
\end{figure}

Figure~\ref{accretion} shows the accretion of gas and of each grain population onto the star and the $5~M_\mathrm{J}$ planet as a function of time. Both 1~cm and 1~mm grains are accreted very efficiently onto the star, about twice as much as the gas, showing their strong inward migration. The complete disappearance of the inner disk for 1~cm grains is confirmed by their accretion rate, which stalls after $\sim80$~orbits. The slightly lower accretion rate of 1~mm grains leaves a small population in the inner disk. The more strongly coupled 100~$\mu$m grains have an accretion rate comparable to that of the gas. Conversely, the accretion onto the planet is very weak for 1~cm grains, which are decoupled and stay in the corotation region on horseshoe orbits (see below), and for 1~mm grains, which quickly migrate out of the gap. The 100~$\mu$m grains take longer to leave the gap and consequently more are accreted by the planet, as is the gas which can still flow through the gap.

The dust behaviour for other planet masses can also be understood from similar $s_\mathrm{opt}$ profiles, which can easily be inferred from the gas surface density profiles in the top right panel of Fig.~\ref{density_profiles} (via $\log s_\mathrm{opt}\ (\mathrm{cm})\simeq\log\Sigma\ (\mathrm{kg\,m}^{-2})-1.4$ from Eq.~(\ref{EqSopt2})). $s_\mathrm{opt}$ does not change much in the outer disk, the shape and density profile of which are therefore the same for all planet masses. For $M_\mathrm{p}=0.1$ to 1~$M_\mathrm{J}$, the gas density and consequently $s_\mathrm{opt}$ do not decrease much in the gap, preventing grains from decoupling and staying in the corotation region. Particles thus tend to evacuate the gap and move towards the pressure maxima at its edges. The inner disk is constantly replenished with grains migrating from the gap, slowing down its accretion onto the star and preventing its disappearance at the end of the simulations for 1~cm grains. Smaller grains have sizes smaller than $s_\mathrm{opt}$ at all times interior to the gap, they therefore migrate very slowly and do not fall onto the star.

\begin{figure}[t]
\resizebox{\hsize}{!}{\includegraphics{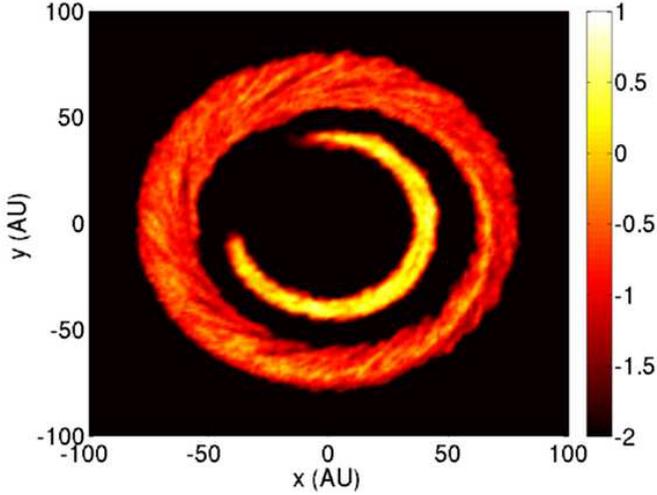}}
\caption{Map of the dust-to-gas ratio, computed as the ratio of volume densities, in the disk midplane for $M_\mathrm{p}=5~M_\mathrm{J}$ and $s=1$~cm (logarithmic scale).}
\label{dust-to-gas}
\end{figure}

The 5~$M_\mathrm{J}$ planet triggers a strong spiral density wave in the gas (see Fig.~\ref{density_maps}) which, incidentally, enhances the gas accretion onto the star \citep[see][]{Laughlin1997}, and is responsible for the lower gas density inside the gap compared to other planet masses (Fig.~\ref{density_profiles}). This spiral wave corresponds to a pressure maximum. Dust particles drift towards its location and the spiral pattern is also seen in the dust phase. The dust-to-gas ratio, shown in Fig.~\ref{dust-to-gas} for 1~cm grains in the disk midplane, is everywhere larger than its initial value of 0.01 and even reaches unity along the spiral wave. (It is still higher in the corotation region where there is little gas left.) This shows that the dust is even more concentrated by the action of the drag than the gas is in the spiral wave, as was found by \citet{Rice2004}.

The variation of the disk thickness with planet mass can be understood in light of the work by \citet{Edgar08}, who establish that there are strong vertical motions of the gas in the spiral arms, and they speculate that this stirs the dust vertically. These motions are naturally stronger for more massive planets, resulting in a thicker dust disk which we see in Fig.~\ref{density_maps}. Because the spiral arms are stronger closer to their launching point, we  also expect vertical motions to be stronger close to the gap edge than in the outer disk and, as a result, the disk is thicker close to the gap for the larger planet masses. The less perturbed disks with lower mass planets have a regular flared shape that is thicker at large radii.

\begin{figure}
\resizebox{\hsize}{!}{\includegraphics{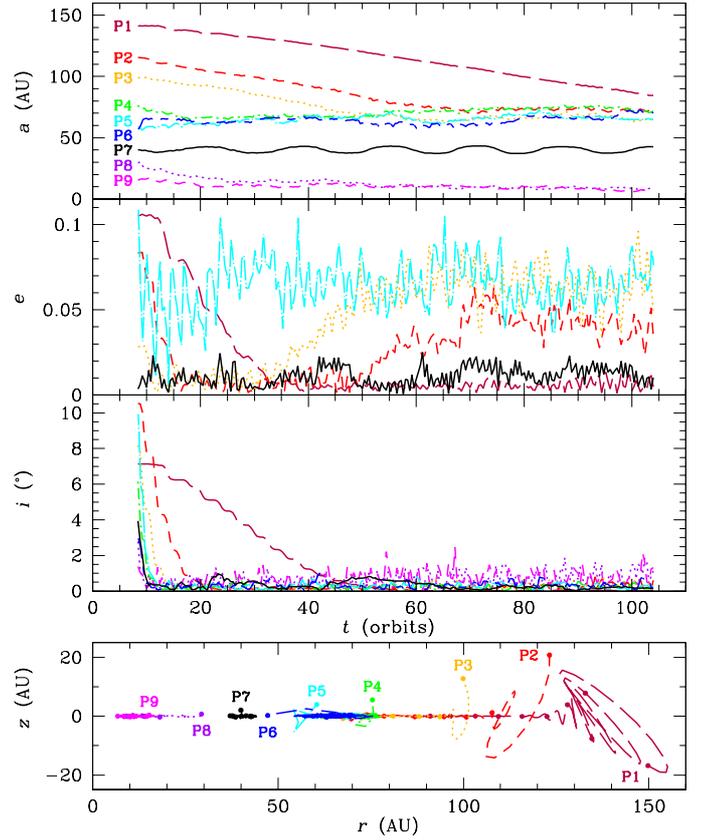}}
\caption{Time evolution of orbital elements $a$, $e$, and $i$ (top panels) and trajectories in the $rz$ plane (bottom) of SPH dust particles initially at various radii in the upper layer of the disk, for the simulation with $M_\mathrm{p}=5~M_\mathrm{J}$ and $s=1$~cm.}
\label{fig_traj_zmax}
\end{figure}

\begin{figure}
\resizebox{\hsize}{!}{\includegraphics{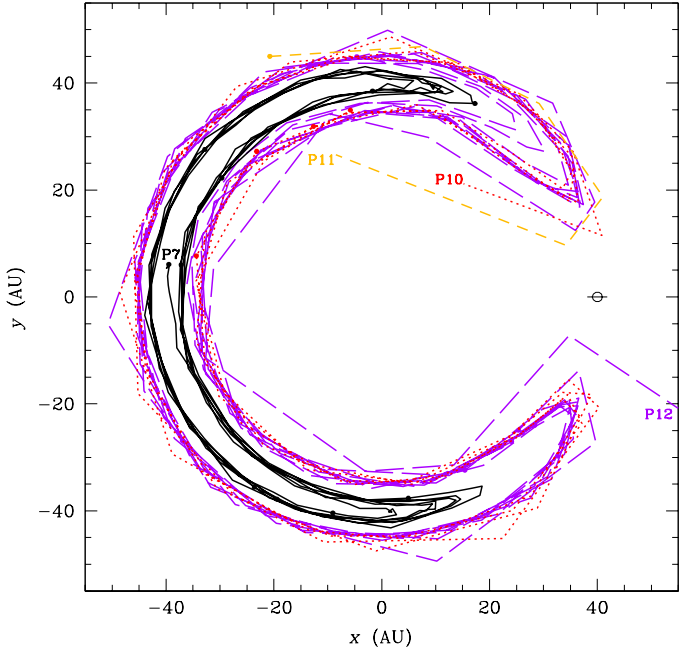}}
\caption{Trajectories in the $xy$ plane, in the planet's reference frame, of SPH dust particles initially in the corotation region, for the simulation with $M_\mathrm{p}=5~M_\mathrm{J}$ and $s=1$~cm. The Saturn-like symbol shows the planet's location.}
\label{fig_traj_corot_xy}
\end{figure}

The Lagrangian nature of SPH can be used to confirm the interpretation of the dust behaviour detailed above by following the trajectories of individual SPH particles. In the following discussion, we focus on the simulation with $M_\mathrm{p}=5~M_\mathrm{J}$ and $s=1$~cm. We select a sample of dust particles initially in the upper layer of the disk at various radii and compute their instantaneous orbital elements (semi-major axis, $a$, eccentricity, $e$, and inclination, $i$) from their positions and velocities at each timestep. Figure~\ref{fig_traj_zmax} plots the time evolution of $a$, $e$, and $i$ as well as the trajectories of these particles in the $rz$ plane. Dots are plotted at regular time intervals (bottom panel of Fig.~\ref{fig_traj_zmax}) to show how fast particles traverse different parts of their trajectories. The particles are initially on inclined elliptical orbits and settle to the midplane while they migrate radially. The efficient damping of their inclination is clearly visible. 
Particles that started in the outer disk (P1, P2, P3, P4) migrate inwards and pile up at the outer gap edge (60--70~AU). Those particles initially between that edge of the gap and the planet (at 40~AU) migrate outwards this time, towards the pressure maximum at the gap edge (P5, P6). Their eccentricities are damped during their migration, but when they reach the gap edge (which the outermost P1 does not achieve by the end of the simulation), $e$ increases again and oscillates about an average value between 0.04 and 0.08. The eccentricities of P4 and P6 behave very similarly to that of P5 but are not plotted to avoid overcrowding the figure. The trajectory of P7, also shown in the $xy$ plane in Fig.~\ref{fig_traj_corot_xy}, is representative of dust grains that started in the vicinity of the planet: they are in corotation and follow horseshoe orbits. Finally, particles initially interior to the corotation region migrate inwards towards the inner gap edge and end up being accreted by the star (P8, P9). Their eccentricities, also not plotted, oscillate between 0 and 0.05.

\begin{figure}
\resizebox{\hsize}{!}{\includegraphics{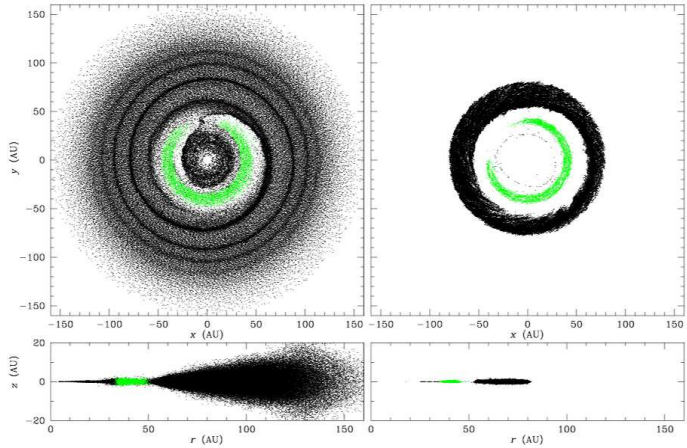}}
\caption{Positions in the $xy$ (top) and $rz$ (bottom) planes of the SPH dust particles at dust injection (left) and at the end of the simulation (right) for $M_\mathrm{p}=5~M_\mathrm{J}$ and $s=1$~cm. Particles ending up in corotation are highlighted in green.}
\label{fig_particles_end_corot}
\end{figure} 

The positions of SPH dust particles at dust injection and at the end of the simulation are plotted in Fig.~\ref{fig_particles_end_corot}. Selecting those that are in the corotation region (in green) at the end of the simulation and looking for their initial positions show that they were already there. This confirms that they decouple very quickly from the gas and do not leave the gap, and that particles initially elsewhere do not migrate into the corotation region.

\begin{figure}
\resizebox{\hsize}{!}{\includegraphics{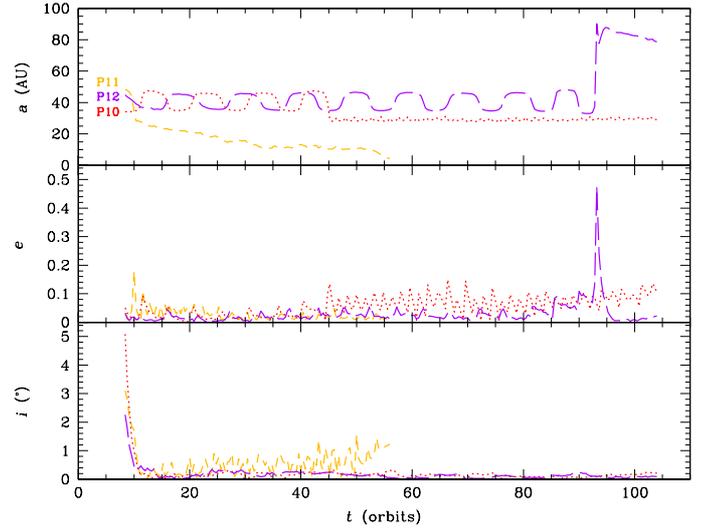}}
\caption{Time evolution of orbital elements of SPH dust particles initially in the corotation region, for the simulation with $M_\mathrm{p}=5~M_\mathrm{J}$ and $s=1$~cm.}
\label{fig_traj_corot_rel_points}
\end{figure}

Some particles initially in the corotation region can pass close to the planet, which then scatters them away. The trajectories of some of these particles up to their ejection are plotted in Fig.~\ref{fig_traj_corot_xy} (the coarseness of the curves is due to the limited frequency of the code data output), while Fig.~\ref{fig_traj_corot_rel_points} shows the time evolution of their orbital elements. P10 is initially on an inclined orbit and settles to the midplane and keeps a horseshoe orbit for some time, then passes close to the planet and is scattered inwards. By that time, there is not enough gas left in the inner disk to make it migrate further ($s_\mathrm{opt}$ has decreased below 1~cm and the particle is decoupled from the gas) and it stays in a marginally perturbed orbit around 30 AU. P11 passes close to the planet early in the evolution, before completing a full horseshoe orbit, and is scattered inwards. It keeps migrating to the inner gap edge, slightly outside 10~AU, where it stays for some time before eventually being accreted by the star together with the rest of the inner disk particles. P12 is scattered outwards with a high eccentricity after a long period on its horseshoe orbit. It then migrates inwards, its semi-major axis and eccentricity being damped by the gas.

\subsection{Understanding the disk asymmetry}
\label{sec:eccentricity}

\begin{figure}
\resizebox{\hsize}{!}{\includegraphics[angle=-90]{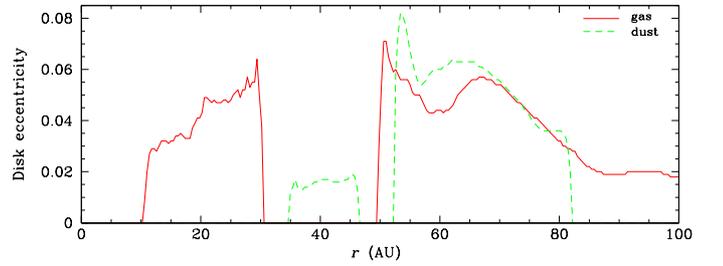}}
\caption{Disk eccentricity as a function of radius for the gas and dust phases for $M_\mathrm{p}=5~M_\mathrm{J}$ and $s=1$~cm.}
\label{disk_e}
\end{figure}

Figures~\ref{density_maps} and \ref{fig_particles_end_corot} show that the gap and outer disk become asymmetric in the dust phase for 1~cm-sized grains with a 5~$M_\mathrm{J}$ planet.
As can clearly be seen in the gas phase of Fig.~\ref{density_maps}, the more massive the planet the stronger the resulting spiral wave. This wave is not a ``physical'' wave in that it comprises specific particles throughout the simulation, but is instead a density wave which moves through the disk with a pattern speed that matches that of the planet's orbit. Thus it is not coherent eccentric orbits of the individual particles, but the density wave itself which creates the asymmetric distribution in the dust phase. The 1~cm particles respond most strongly to the pressure maximum in the density wave and therefore we see a stronger disk asymmetry. This is enhanced by the fact that the disk of 1~cm-sized grains is more compact, making the structure more obvious.

\cite{Kley2006} studied gas-only disks with an embedded planet on a circular orbit and observed a transition from a circular to an eccentric disk for planet masses larger than 5.25~$M_\mathrm{J}$ in an $\alpha_\mathrm{SS}=0.01$ disk. They found that the disk is turned eccentric by the excitation of an $m=2$ spiral wave at the outer 1:3 Lindblad resonance. The asymmetry in our dust disk cannot have the same origin because our planet masses are smaller than that necessary for an eccentric disk, and because our simulations are run for 100 orbits, which is much shorter than the viscous time scale necessary for eccentricity growth. We computed the disk eccentricity, defined as the average of the instantaneous eccentricity of all particles at a given radius and which \cite{Kley2006} used as a diagnosis tool, and plot it for gas and dust in Fig.~\ref{disk_e}. For both phases, the disk eccentricity reaches a maximum of only about 0.08, similarly to their non-eccentric disks (see their Fig.~2). Finally, we do not see any asymmetry in our gas disk. The one we observe in the dust disk is therefore specific to the dust dynamics in the presence of a spiral wave in the gas.

As expected from the analytical study of \citet{Marzari00}, we do not observe periastron alignment of our dust particles on neighbouring orbits because our planet is not eccentric. However, contrary to \citet{Ciecielag07} we do not see a correlation between periastron longitude and eccentricity. This is likely due to the fact that our disk perturbation is much weaker due to the lower mass of our secondary compared to theirs.

The disk asymmetry has implications for observations. \citet{TA01} suggested that consistent detections of planets through structures in the dust layer would be more convincing if asymmetric. They showed that it was possible to build axisymmetric dusty rings and gaps without the presence of a planet. Here we show that even a planet on a circular orbit can lead to an asymmetric dust structure, and for a grain size that will be observable by ALMA.

\subsection{Understanding torques}
\label{sec:torques}

\begin{figure}
\resizebox{\hsize}{!}{\includegraphics[angle=-90]{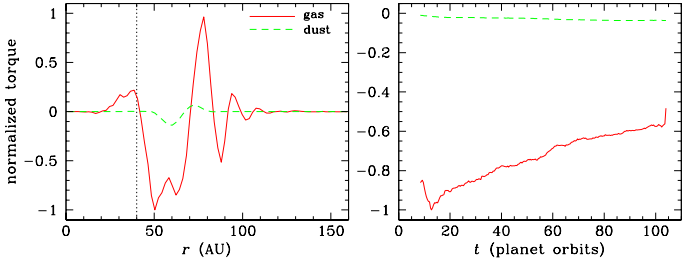}}
\caption{Torques due to the gas and dust phases on the planet in the simulation with $M_\mathrm{p}=5~M_\mathrm{J}$ and $s=1$~cm. Left: normalized torques as a function of the radius in the disk at the end of the simulation. The planet's orbit is marked by the dotted line. Right: normalized total torque on the planet as a function of time.}
\label{torques}
\end{figure}

In \citetalias{Fouchet07} we suggested that the settled dust may have a non-negligible effect on planet migration. While the dust content is only 1\% in mass of that of the gas, once it has settled to the midplane where the planet orbits, it could exert a rather strong torque on it. We therefore computed torques exerted by the dust and compared it to those exerted by the gas in Fig.~\ref{torques} for our simulation with $M_\mathrm{p}=5~M_\mathrm{J}$ and $s=1$~cm. In the left panel, we plot the torques as a function of radius at the end of the simulation. The torque exerted by the gas shows a positive component inside the orbit of the planet and a negative one outside, a well-known feature of planet gaps. The oscillations at larger radii are due to the spiral wave. The inner component is substantially weaker than the outer one because the inner disk is partly depleted. The torque exerted by the dust phase has the same global shape except that the inner component vanishes because the inner dust disk is empty. It is an order of magnitude smaller than that due to the gas because the densities also differ by an order of magnitude. For 1~mm and 100~$\mu$m grains, the dust-to-gas ratio is even smaller, resulting in an even weaker dust torque (not shown). We therefore do not expect dust structures to have a direct dynamical influence on planet migration. 

In the right panel of Fig.~\ref{torques}, we plot the total torque as a function of time. Here we can see that the absolute value of the torque exerted by the gas steadily decreases while that of the dust slowly increases. This is expected given that the gap in the gas phase is being carved until it reaches a steady state after several hundreds of orbits. Thus, the gas in the Lindblad resonances (which are inside the Hill radius) is depleted and the Lindblad torque decreases. The dust, on the other hand, is piling up at the outer gap edge, which results in the increase of the outer Lindblad torque. Eventually, as the system reaches a steady state, the increase in the Lindblad torque exerted by the dust phase will stop as well. Perhaps, when the stationary case is reached, will dust have a direct dynamical influence on the planet.

\subsection{Effect of backreaction of dust on gas}
\label{sec:BR}

\begin{figure}
\resizebox{\hsize}{!}{\includegraphics[angle=-90]{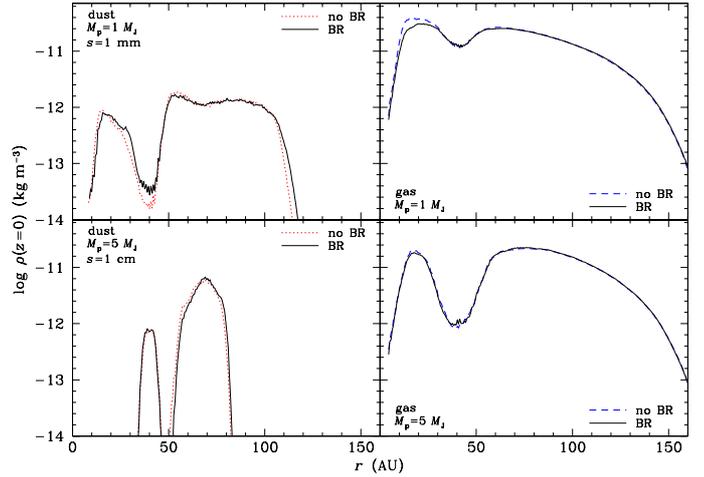}}
\caption{Azimuthally averaged midplane density profiles after 104 planetary orbits for $M_\mathrm{p}=1~M_\mathrm{J}$ and $s=1$~mm (top) or $M_\mathrm{p}=5~M_\mathrm{J}$ and $s=1$~cm (bottom), for simulations with or without backreaction (BR) of dust on gas. Left: dust; right: gas.}
\label{fig_BR}
\end{figure}

The simulations presented so far are all run with no backreaction of the dust phase on the gas. To see whether this simplification is valid, we plot in Fig.~\ref{fig_BR} the gas and dust volume densities for simulations with and without backreaction of dust on the gas, for two configurations: ($M_\mathrm{p}=1~M_\mathrm{J}, s=1$~mm) and ($M_\mathrm{p}=5~M_\mathrm{J}, s=1$~cm).

In the (1~$M_\mathrm{J}$,1~mm) case, the effect of dust on the gas is mostly visible in the inner disk, where it drags the gas along with it towards the central star, decreasing the gas density. Indeed, after 104 orbits, the number of gas SPH particles accreted by the star is about 10\% larger with than without backreaction. In the dust, when backreaction is included, we see that the depth of the gap and the height of the outer edge are smaller and the outer disk radius is larger. This is because the backreaction means that the dust tends to drag the gas along with it, which effectively reduces the gas drag on the dust. Therefore, its effects on the dust distribution (radial migration, gap creation, pileup at the outer edge) which we have discussed extensively are weakened. This means that the dust depletion in the gap is not as strong, the dust pileup at the gap edges is weakened and the radial migration in the outer disk is retarded.

The effects for the (5~$M_\mathrm{J}$,1~cm) case are the same, and one may expect them to be more important for a larger planet mass causing stronger pressure gradients and for the grain size causing the largest dust-to-gas ratio. However, they are harder to see, for several reasons. The effect on the gas in the inner disk is smaller because all the dust there has been accreted onto the star well before the end of the simulation, and the time during which the backreaction of dust on gas can act is much smaller than in the (1~$M_\mathrm{J}$,1~mm) case. The effect of backreaction on the dust in the inner disk cannot be evaluated since the dust has disappeared there. In the gap, dust grains decouple from the gas very early on and stay on horseshoe orbits, their distribution is no longer affected by gas drag, with or without backreaction, and both curves are very similar. With backreaction, the height of the outer edge in the dust is slightly reduced, as for the (1~$M_\mathrm{J}$,1~mm) case. Finally, the dust disk outer radius is also slightly larger with backreaction, but this is harder to see due to the steeper profile.

The overall effects of backreaction in the simulations presented in this paper are quite small, however, which validates our assumption of no backreaction for the simulations presented in the rest of this study. We do not investigate in this work the potential development of the streaming instability, which, in the presence of backreaction of dust on gas, can lead to density enhancements and clumping in the dust layer \citep[for more details, see][]{Youdin05,Youdin07}.

\subsection{Putting our work in context}
\label{sec:context}

Over the past few years there has been a wealth of literature on the effects an embedded planet has on the gas and dust grains of various sizes in protoplanetary disks. \citet{Edgar08} studied the vertical structure of spiral arms raised by an embedded planet, suggesting that the strong gas vertical motions should make the dust layer slightly thicker close to the gap (or at least over the radial extent of the spiral wave). Our simulations of gas and dust disks confirm their expectations as discussed in Sect.~\ref{sec:gap}, which can have implications for the visibility of the gap.

\citet{PM04,PM06} studied the effects of low mass planets ($M_\mathrm{p}=0.01$ to 0.5~$M_\mathrm{J}$) on strongly coupled grains (1~mm in size but with a denser and more compact nebula than in our study) in vertically integrated 2D disks. They were therefore unable to follow the vertical settling of dust and to investigate the variation of the disk thickness with grain size and the interplay between vertical and radial migration as we do. Because their grains were strongly coupled to the gas, the structures in their disks where mostly due to the planet perturbation on the gas. Decoupling between gas and dust happened at shock fronts in the gas where it was decelerated while dust, with no internal pressure, was not. As a result, the flux of dust grains inside the spiral arms was larger than that of the gas and, because spiral arms deflect particles away from the planet, the gap in the dust layer became deeper than that in the gas phase. In our case, planets are massive enough to directly deflect particles, which subsequently stay out of the gap (see Sect.~\ref{sec:gap}). Indeed, their eccentricity acquired during the close encounter with the planet is rapidly damped by the gas and they slowly migrate towards the outer gap edge, if deflected outwards (like particle P12 in Fig.~\ref{fig_traj_corot_rel_points}).

Additionally, while \citet{PM04,PM06} observe resonant trapping (through an indirect mechanism), we do not (this has been discussed in \citetalias{Fouchet07}). This difference can be traced back to the fact that our dust is marginally decoupled while theirs is strongly coupled despite the fact that we both study 1~mm grains (again our nebula parameters differ). As a result, they expect multi-ringed structures which could be observationally difficult to disentangle from a multi-planet system. We, on the other hand, see structures that can only be related to a single planet. We expect that smaller grains would behave similarly to what was found by \citet{PM04,PM06} but, in the submillimetre wavelength range, those smaller grains do not contribute much. We thus conclude that, for the nebula parameters we chose, there will be no ambiguity between single or multiple planet systems.

Finally, as well as potentially exerting a torque on the planet, changes in the dust density will affect the dust opacity, which strongly influences the temperature structure around the planet. We study grains in a size range where their opacity is not dominant, but due to collisional shattering, they most likely give rise to a population of smaller grains with large opacities who will tend, at least to first approximation, to follow the spatial distribution of their larger parent grains. Recent studies using radiative transfer in the flux-limited diffusion approximation \citep{PM06b,PM08} show that the temperature structure in the disk and specifically around the planet has a strong effect on the migration rate, but also on the migration direction. \citet{Hasegawa10a} have shown that dust settling and the presence of a dead zone have an effect on the temperature distribution. They note the formation of a dust wall at the edge of the dead zone that is directly illuminated by the star and see a local positive temperature gradient in front of the dust wall. They demonstrate that it has an effect on the migration of small planets in a forthcoming paper \citep{Hasegawa10b}. It is however beyond the scope of our paper to study the variations in opacity and subsequent effect on planet dynamics.

\section{Conclusions}
\label{sec:conclusion}

We have run 3D, two-phase (gas and dust) SPH simulations of a typical CTTS disk of mass 0.02~$M_{\sun}$ with an embedded giant planet on a circular orbit at 40~AU. We vary the grain size (100~$\mu$m, 1~mm, 1~cm) and the planet mass (0.1, 0.5, 1, 5~$M_\mathrm{J}$) and study the formation of the planetary gap. We confirm that gap opening is stronger in the settled dust layer than in the flared gas disk. Gaps are deeper and wider for (1) larger, more efficiently settled grains and (2) more massive planets. Larger planet masses are required to open a gap in the gas phase than in the dust: while a 0.5~$M_\mathrm{J}$ planet only slightly affects the gas phase, it carves a deep gap in the dust. For the most massive 5~$M_\mathrm{J}$ planet, 1~cm grains remain trapped in corotation with the planet while their distribution in the outer disk shows an asymmetric structure, even though the planet's orbit is circular. We find that this is not caused by the periastron alignment of a coherent set of orbits but rather by the pile-up of dust in the pressure maximum of the gas phase caused by the spiral density wave triggered by the planet. This global asymmetry does not appear for less massive planets because the spiral perturbation is substantially weaker.

The variety of structures that we obtain in the dust phase for various grain sizes and planet masses has implications on the appearance of protoplanetary disks at (sub)millimetre wavelengths and show how important it is to go beyond the gas-only disk description proposed by, e.g., \citet{W05}. The observability of these disks with ALMA is the subject of a forthcoming companion paper.

\begin{acknowledgements}
This research was partially supported by the Programme National de Physique
Stellaire and the Programme National de Plan\'etologie of CNRS/INSU, France,
the Agence Nationale de la Recherche (ANR) of France through contract
ANR-07-BLAN-0221, the Swinburne University Research Development Grant Scheme,
and the Australia-France co-operation fund in Astronomy (AFCOP).
Simulations presented in this work were run on the Swinburne
Supercomputer\footnote{\tt http://astronomy.swin.edu.au/supercomputing/}
and at the Service Commun de Calcul Intensif (SCCI) de l'Observatoire de
Grenoble, France.
Images in Fig.~\ref{density_maps} were made with SPLASH \citep{Price2007}.
We thank the anonymous referee whose comments have greatly improved this paper.
\end{acknowledgements}

\bibliographystyle{aa}
\bibliography{ms}

\end{document}